\documentstyle[psfig,epsf,conf-X]{article}
\begin{document} 
\small
\heading{%
%
Where are the type 2 AGNs ?
}
\par\medskip\noindent
\author{%
Marco Salvati$^{1}$, Roberto Maiolino$^{1}$
}
\address{%
Osservatorio Astrofisico di Arcetri, Largo E.~Fermi~5, I-50125~Firenze, Italy}

\begin{abstract}
Most of the energy of the X-ray background is contributed
by sources with a hard X-ray spectrum, harder than the
spectrum observed in nearby type 1 AGNs. The hard 
background contributors are conventionally identified with
type 2 AGNs, obscured by circumnuclear matter both in the
optical and in the soft X rays. However, preliminary results
of identification programs seem to find systematically few
type 2 AGNs in samples selected with obscuration insensitive
criteria. We discuss this issue, and assess the various
scenarios which are being proposed to solve the discrepancy.
\end{abstract}
%
%


\section{The distribution of N$_H$.}

Past hard X-ray surveys of Sy2s were strongly biased in favor of X-ray bright
sources (according to all-sky surveys), which tend to be the least absorbed
ones. We used BeppoSAX to observe an [OIII]-selected sample of 
previously unobserved, X-ray weak Sy2s. We found (\cite{salv77}, \cite{maio98})
a large fraction of Compton thick objects, a result which confirms the bias 
against heavily obscured systems in previous surveys. We then merged
all the available hard X-ray observations of Sy2s and extracted a complete
subsample limited in {\it intrinsic} (i.e. unabsorbed) flux as
inferred from the [OIII] narrow emission line \cite{risa99}.
This subsample is composed of 45 objects and the corresponding N$_H$
distribution, shown in Fig.~1, can be considered the best approximation
to the true distribution allowed by the available data.
The most interesting result is that
this distribution is significantly shifted toward large columns with respect
to past estimates: most ($\sim$75\%) of the Sy2s are heavily obscured
($\rm N_H > 10^{23} cm^{-2}$) and about half are Compton thick. 
We have completed the N$_H$ distribution by adding the
Sy1s. Their number relative to the Sy2s is taken from Maiolino \& Rieke
\cite{mr95}, the relative populations of the N$_H$ bins (only cold
absorption is considered) are taken from literature data pertaining to
the Sy1s of the Piccinotti sample \cite{picci}: this is a hard X-ray 
selected sample, and should not miss X-ray absorbed Sy1s unless they are
Compton thick. We note a clear dichotomy between X-ray absorbed and
unabsorbed sources, and a close correspondence between the X-ray and the
optical classification, at least for these local, low luminosity AGNs.

\begin{figure}[!ht]
\parbox{0.499\linewidth}{
\centerline{
\psfig{file=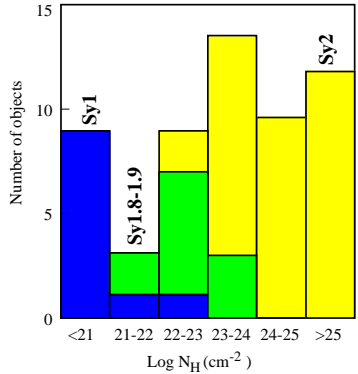,width=0.8\linewidth}
}}
\parbox{0.49\linewidth}{
\small {\caption{ 
The distribution of absorbing column densities among Seyfert galaxies.
From \cite{risa99}, with modifications.}}
}
\end{figure}

\section{The relation between optical and X-ray absorption.}

If the obscuring torus has the Galactic gas-to-dust ratio,
and the dust has the Galactic extinction curve, then
the nuclear region of X-ray absorbed AGNs should suffer a visual 
extinction that is related to
the gas column density by the formula A$_V=4.5\times 10^{-22}$ N$_H$
(cm$^{-2}$). In general this is not the case:  A$_V$ is lower than what
expected from the N$_H$ measured in the X-rays. This was first pointed out by 
Maccacaro et al. \cite{macca82}, and is also required to fit the far-IR
spectrum of AGNs \cite{grana97}. We have collected a sample of Seyferts
which exhibit at the same time X-ray (cold) absorption and optical or IR 
broad lines, not completely suppressed by the intervening matter. 
By assuming the standard extinction curve, and neglecting collisional or 
radiative transport effects, we can derive an estimate of the
visual extinction. A preliminary result is shown in Fig.~2, where the
distribution of A$_V$/N$_H$ relative to the Galactic value is reported.
With the exception of three low luminosity AGNs ($\rm L_X < 10^{41.5} erg
~s^{-1}$, shaded in the histogram),
whose nuclear physics might be intrinsically different \cite{ho99},
most AGNs are characterized by a deficit of dust absorption, in
agreement with early claims. At higher, quasar-like luminosities there are
even more extreme examples of this effect:
objects that, although absorbed in the X rays, do not show significant dust
absorption in the optical and appear as type 1, broad line AGNs
have been recently discovered in hard X-ray and radio surveys (\cite{ree97},
\cite{sam99}, \cite{aki99}, \cite{fiore}). 

The origin of this effect is not clear. An obvious explanation is that the
dust-to-gas ratio is much lower than Galactic. However, large amounts of
dust must be present in the vicinity of active nuclei,
as inferred from their powerful IR emission. Alternatively, most of the
cold X-ray absorption observed in these objects might be caused by broad line
clouds intervening along our line of sight. However, if the low A$_V$/N$_H$
is a property common to most AGNs, this interpretation would require a very
large covering factor for the BLR whereas this is
estimated to be only $\sim$10\%. Dust in the internal region of the torus
might be destroyed by the powerful nuclear radiation field, thus making the
effective dust column significantly lower than the gas column.
However, inside the sublimation radius a huge HII region would be created,
with a covering factor much larger than observed 
(see also Netzer \& Laor \cite{nl93}). Another interesting
possibility is that the dust extinction curve is much flatter than the
Galactic one. The high density of the gas
in the circumnuclear region of AGNs
is likely to favor the growth of large grains which, in turn, should flatten
the extinction curve. This effect is directly observed in the dense clouds of
our Galaxy \cite{dra95}. Large grains have also been proposed to explain the
lack of the 10$\mu$m silicate emission feature in the mid-IR
spectra of AGNs \cite{ld93}. 
Within the context of the optical versus X-ray absorption, the effect
of a flat extinction curve is twofold: 1) given the same dust mass, the
effective visual extinction is lower, and 2) the broad lines ratio gives a
deceivingly low measure of the extinction. A more thorough discussion
of the whole issue is given in Maiolino et al. (in prep.).

\begin{figure}[!ht]
\parbox{0.499\linewidth}{
\centerline{
\psfig{file=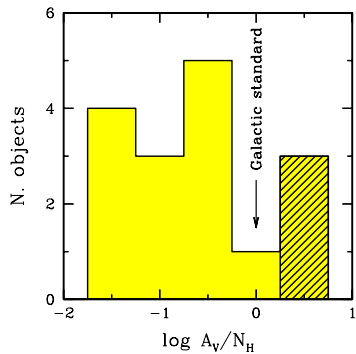,width=0.8\linewidth}
}}
\parbox{0.49\linewidth}{
\small{ \caption{
Distribution of the A$_V$/N$_H$ ratio, relative to the Galactic
value, for a sample of absorbed Seyferts. Three low luminosity sources 
($\rm L_X < 10^{41.5} erg ~s^{-1}$) are indicated with a shaded bin.}}
}
\end{figure}

At the other extreme an increasing number of obscured powerful AGNs has
been discovered by means of hard X-ray observations in galaxies which are
optically classified as starburst or LINER. Probably, in these objects the
obscuring medium hides the NLR as well as the BLR. 
Alternatively, the nuclear ionizing
source might be completely embedded and obscured in all directions. The
most spectacular case is the nearby (4 Mpc) edge-on galaxy NGC4945, whose
nucleus is one of the brightest AGNs at 100 keV \cite{don96}, but
is heavily obscured in all directions. Indeed, optical
to mid-IR observations were unable to detect any trace of AGN
activity (\cite{maio99}, \cite{mar99}).

\section{Implications for the X-ray background.}

Obscured AGNs are thought to be a key ingredient of the hard X-ray
background (XRB, \cite{sw89}, \cite{com95}), and the
distribution of N$_H$ represents the main set of free parameters in the
XRB synthesis models. The N$_H$ distribution presented above can be
used to freeze this set of parameters, under the assumption that it
does not evolve with redshift. A detailed model that takes into 
account this constraint is presented in Gilli et al. \cite{gil99}: although
the shape of the XRB is well reproduced, the number counts
are not, and suggest a larger number of X-ray absorbed AGNs
as compared with the local Universe.

The deficiency of dust absorption with respect to the X-ray absorption,
especially at high luminosities,
implies a possible mismatch between the optical and the X-ray
classification of the sources contributing to the hard X-ray background.
In particular, some of the type 2 QSOs, which many models
assume to make most of the hard XRB, could be optically ``masked'' as type~1 
QSOs and be already present in optical surveys. On the other hand, the
existence of heavily obscured AGNs in powerful IR galaxies which optically 
are {\it not at all} classified as AGNs suggests that a fraction of ULIRGs
and SCUBA sources might host some of the type 2 QSOs which are needed for
the hard XRB.

\begin{acknowledgements}
Many of the new results presented in this paper were obtained in collaboration
with R. Gilli, A. Marconi, and G. Risaliti.
This work was partially supported by the Italian Ministry for University and
Research (MURST) through the grant Cofin-98-02-32.
\end{acknowledgements}

\vfill
\end{document}